\documentclass[preprint,
superscriptaddress,
groupedaddress,
 amsmath,amssymb,
 aps,
]{revtex4-2}
\usepackage{graphicx}
\usepackage{dcolumn}
\usepackage{bm}
\usepackage{mathtools}
\usepackage{xcolor}
\DeclarePairedDelimiter\abs{\lvert}{\rvert}
\DeclareMathOperator{\sech}{sech}
\usepackage{hyperref}

\begin{document}
\preprint{APS/123-QED}
\title{Magnetic Solitons due to interfacial chiral interactions}

\author{Paula Mellado}
\affiliation{
Facultad de Ingenieria y Ciencias, 
Universidad Adolfo Ibáñez,
	Santiago, Chile
}
\author{Ignacio Tapia}
\affiliation{Facultad de Ciencias, Departamento de Fisica, Universidad de Chile, 
Casilla 653, Santiago, Chile}
\date{\today}
\begin{abstract}
We study solitons in a zig-zag lattice of magnetic dipoles. The lattice comprises two sublattices of parallel chains with magnetic dipoles at their vertices. Due to orthogonal easy planes of rotation for dipoles belonging to different sublattices, the total dipolar energy of this system is separable into a sum of symmetric and chiral long-ranged interactions between the magnets where the last takes the form of  Dzyaloshinskii-Moriya coupling.  For a specific range of values of the offset between sublattices, the dipoles realize an equilibrium magnetic state in the lattice plane, consisting of  one chain settled in an antiferromagnetic parallel configuration and the other in a collinear ferromagnetic fashion. If the offset grows beyond this value, the internal Dzyaloshinskii-Moriya field stabilizes two Bloch domain walls at the edges of the antiferromagnetic chain. The dynamics of these solitons is studied by deriving the long-wavelength lagrangian density for the easy axis antiferromagnet. We find that the chiral couplings between sublattices give rise to an effective magnetic field that stabilizes the solitons in the antiferromagnet. When the chains displace respect to each other, an emergent Lorentz force accelerates the domain walls along the lattice. 
\end{abstract}
\maketitle
\section{\label{sec:intro}Introduction}
Magnetic solitons \cite{kosevich1990magnetic}, non-linear excitations in magnetic systems,  are known for their potential to encode and manipulate information efficiently, and with minimum dissipation, \cite{ochoa2018spin}. To that purpose, once they materialize in a system,  it is crucial to create the conditions for their stability \cite{karashtin2015instability} and to apply the rightful force for their propulsion. Magnetic solitons can exist in ferromagnetic and antiferromagnetic materials \cite{ivanov1980dynamics}. Though both scenarios are attractive from fundamental and technological points of view \cite{baltz2018antiferromagnetic,turov1996symmetry}, in practical matters, antiferromagnetic systems offer some advantages over their ferromagnetic counterparts like faster dynamics, exchange enhancement of the order parameters \cite{jungwirth2016antiferromagnetic}, and absence of stray fields, which allows resonance frequencies in the terahertz range \cite{galkina2018dynamic}. Nevertheless, studying the dynamics of antiferromagnetic solitons presents new challenges due to the absence of net magnetization and the emergence of Lorentz forces analogs of the gyroscopic forces in ferromagnets \cite{dasgupta2017gauge}. 
Though it has been shown that magnetic field alone, or the adiabatic spin torque from an applied electric current, cannot propel solitons in an antiferromagnet \cite{dasgupta2017gauge,pan2018driving}, there is evidence supporting that a hybrid strategy could accelerate antiferromagnetic solitons by for instance combining magnetic fields with the torque due to magnetic current from magnons \cite{kim2014propulsion,dasgupta2018energy} or an adiabatic spin current \cite{shen2020driving}.

\emph{Summary of  results.} 
Aimed to identify new venues for realizing magnetic solitons in antiferromagnetic materials and find new schemes for their dynamics,  we study  a zig-zag lattice of magnetic dipoles. The lattice extends along the $\hat{x}$ axis and comprises two parallel sublattices with an offset along $\hat{y}$, each with magnetic dipoles at their vertices. Dipoles in different sublattices have perpendicular easy planes of rotation and couple via the long-ranged dipolar interaction.

In the first part of the paper we examine the magnetization dynamics of the zig-zag lattice when the offset between sublattices is tuned.  For this purpose we solve the set of coupled equations of motion for the angular rotation of each dipole using molecular dynamics simulations. The angular dynamics is caused by the internal torques resulting from the dipolar interactions between the magnets. For a given dipole, the internal torques propelling its rotation can be changed if the offset between sublattices is varied.
We find that depending on the offset, the system relaxes into four possible magnetic configurations. Here we focus our analysis in a range of offsets that allow the lattice to settle in the planar phase:  with one sublattice relaxed in a collinear magnetic configuration where dipoles order along the lattice, and the other in a state where dipoles settle in a parallel antiferromagnetic arrangement orthogonal to the lattice. While the lattice is in the planar phase, as the offset is increased, two Bloch domain walls (DW) arise at the edges of the antiferromagnetic chain while the angular positions of the ferromagnetic sublattice remain motionless. As the offset increases further, the two DW move from the edges of the lattice until they meet and annihilate at its center. As the offset is increased even further, the system returns to the planar phase.

In the second part of the paper we examine a continuum model that allows  unveiling the internal fields responsible for the DW birth  and identifying the forces that promote their translations. We find that solitons are stabilized by an emergent magnetic field product of symmetric and chiral internal fields rooted in the inter-sublattice dipolar interactions. We show that dynamics of such solitons along the zig-zag lattice is due to the symmetric coupling between sublattices and is induced by an internal electromagnetic force that emerges along the lattice when one sublattice moves with respect to the other.

The rest of the paper is organized as follows. Section \ref{sec:model} describes the model, examines the magnetic equilibrium states of the zig-zag lattice as a function of the offset and identifies the inter and intra-chain symmetric and chiral long-ranged couplings between dipoles.  In order to understand the dynamics of the Bloch domain walls arose in the planar phase of the lattice, in section \ref{sec:lagrangian} we write the field theory for the antiferromagnetic chain. By describing solitons in terms of collective coordinates,  we identify the electromagnetic force on the soliton in regard to the internal dipolar couplings. This allows the formulation of a new strategy for propelling solitons in antiferromagnetic chains. Section \ref{sec:conclusion} is devoted to discussion and concluding remarks. Appendix \ref{sec:sim} and Appendix \ref{sec:ana} show respectively details of the numerical simulations and analytical calculations.
\begin{figure*}
\includegraphics[width=\textwidth]{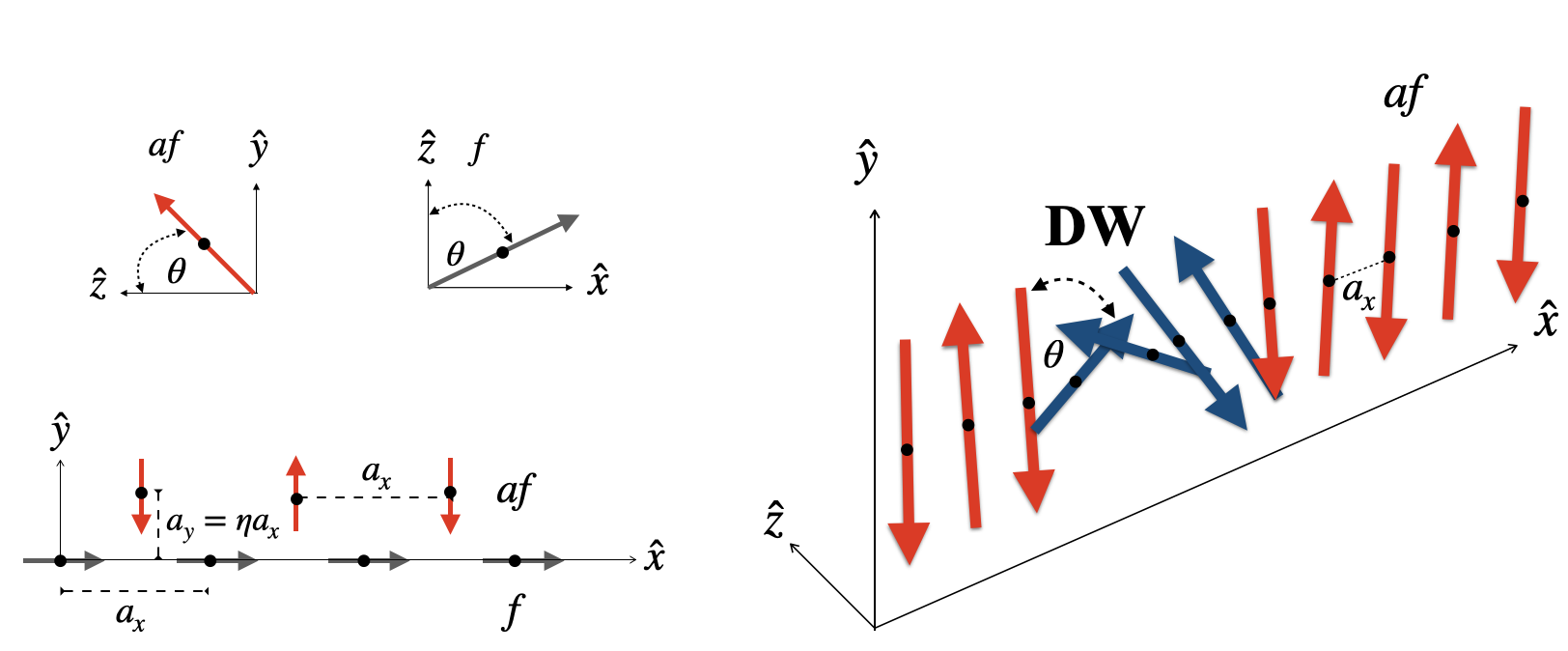}
\caption{Left: The zig-zag lattice extends along the $\hat{x}$ axis and consists of dipoles located in two parallel chains denoted $af$ and $f$ and shown in red and grey, respectively. $af$ and $f$ have lattice constant $a_x$ and are locate at a distance $a_y$ apart along the axis $\hat{y}$. Dipoles in $af$ are at the middle point between two nearest neighbor dipoles in $f$. Dipoles in $af$ and $f$ rotate with angle $\theta$ in perpendicular planes. Right: A Bloch domain wall in \emph{af}, like the ones found at the in the simulations, is illustrated in blue.} 
\label{f1}
\end{figure*}
\begin{figure*}
\includegraphics[width=0.8\textwidth]{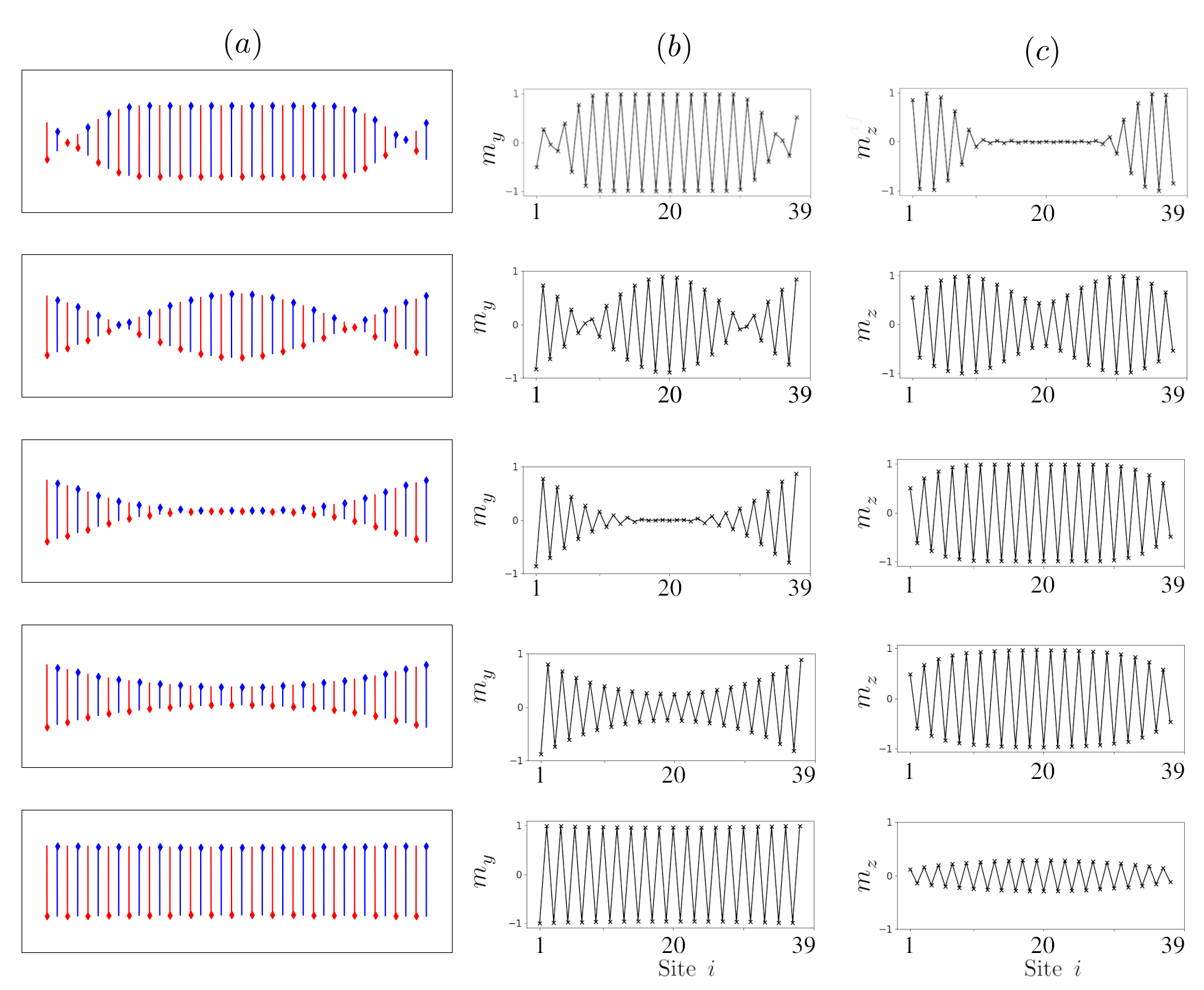}
\caption{Results from molecular dynamics simulations of the zig-zag lattice with $n=39$ dipoles in \emph{f} and 38 dipoles in \emph{af}. From top to bottom, $\eta=1.3, 1.36, 1.38, 1.94$. (a) Equilibrium magnetic configurations of dipoles in sublattice $af$ projected in the plane $\hat{x}-\hat{y}$. In the two upper rows, two Bloch DWs are apparent close to the edges. As $\eta$ grows, the DWs move toward the center of the chain until they meet (third row) and restore the sublattice to its original ground state (lower row). (b) Magnetization along the $\hat{y}$ direction, $m_y$, of dipoles in (a). (c) Magnetization along the $\hat{z}$ direction, $m_z$ of dipoles in (a).} 
\label{f2}
\end{figure*}
\begin{figure}
\includegraphics[width=0.7\columnwidth]{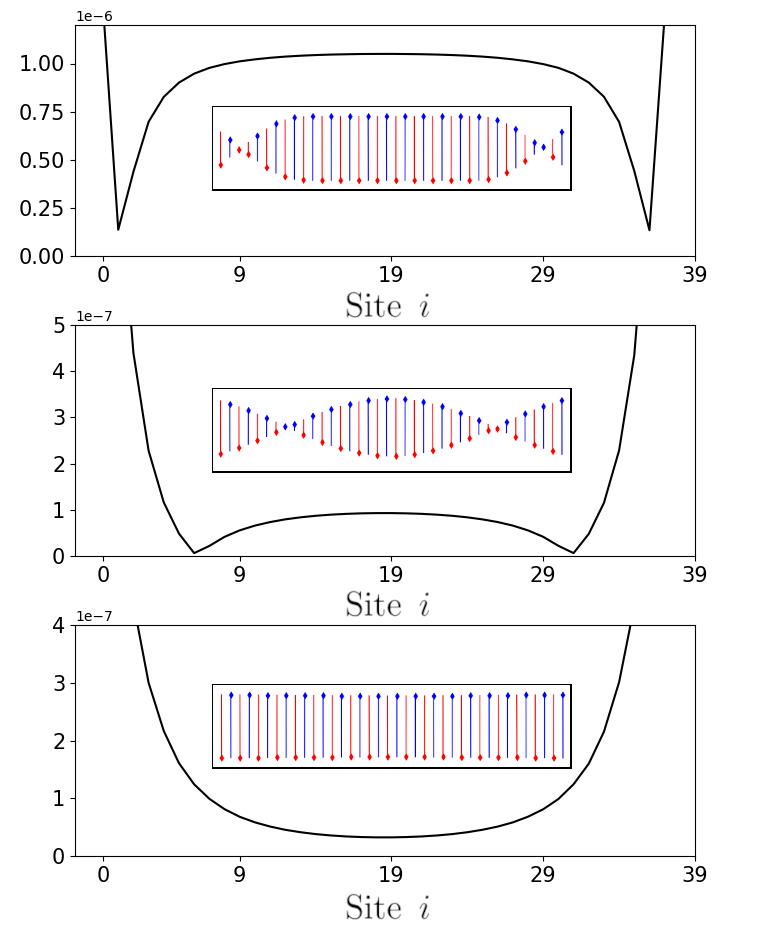}
\caption{Energy landscape along \emph{af} due to sublattice \emph{f} for the system of Fig.\ref{f2}. From top to bottom, $\eta=0.98, 1.30, 1.94$. In the upper row, the two minima at the edges of \emph{af} coincide with the positions of the solitons. As $\eta$ grows, the potential wells become shallower and move toward the center of \emph{af} (middle row). The lower row corresponds to the landscape of \emph{af} when it has returned to the ground state.} 
\label{f3}
\end{figure}
\section{\label{sec:model}Model}
The magnetic dipolar energy for dipoles in the zig-zag lattice reads 
\begin{equation}
    U_d=\frac{g}{2}\sum_{\alpha,\beta}\sum_{i,k}\frac{\hat{\bm m}_i^\alpha\cdot\hat{\bm m}_k^\beta - 3(\hat{\bm m}_i^\alpha\cdot\hat{\bm{r}}_{ik}^{\alpha\beta})(\hat{\bm m}_k^\beta\cdot\hat{\bm{r}}_{ik}^{\alpha\beta})}{|{\bm r}_i^\alpha-{\bm r}_k^\beta|^3}
    \label{eq1}
\end{equation}
where $\hat {\bm{r}}_{ik}^{\alpha\beta}= \frac{({\bm r}_i^\alpha -{\bm r}_k^\beta )}{|{\bm r}_i^\alpha -{\bm r}_k^\beta |}$ denotes the unit vector joining dipole $\hat{\bm m}_i^\alpha$ at site $i$ and sublattice $\alpha$ and dipole $\hat{\bm m}_k^\beta$ at site $k$ and sublattice $\beta$. $g =\frac{\mu_0 m_0^2}{4\pi a_x^3}$ sets an energy scale \cite{mellado2022intrinsic} and contains the physical parameters of the system, such as $a_x$, the lattice constant along the $x$ direction, $\mu_0$, the magnetic permeability, and $m_0$, the intensity of the magnetic moments. The zig-zag lattice can be seen as a one dimenssional lattice with a two point basis, or sublattices. Dipoles are fixed at the vertices of  sublattices \emph{f} and \emph{af} as shown in the left panel of Fig.\ref{f1}. The sublattices are coplanar parallel chains that extend along the $\hat{x}$ direction and are set apart along $\hat{y}$ by distance $a_y=\eta a_x$. The sites of sublattice \emph{af} are set at the middle point between two nearest neighbor sites of sublattice \emph{f} as shown in Fig.\ref{f1} (left). \emph{f} contains $n$ dipoles at its sites and \emph{af} has $n-1$. Dipoles in \emph{f} rotate in terms of a polar angle $\theta$ in the plane $\hat{x}-\hat{z}$, that is respect to a local axis along $\hat{y}$, fixed to their center. Dipoles in \emph{af}, on the other hand, rotate in terms of $\theta$ in the  $\hat{y}-\hat{z}$ plane, that is, respect to a local axis along $\hat{x}$ (see Fig.\ref{f1}). All dipoles of the zig-zag lattice have fixed  angular positions respect to the azimuthal angle $\varphi^\alpha: \varphi^\emph{f}=0$ and $\varphi^\emph{af}=\frac{\pi}{2}$. Hereafter the magnetic moments are normalized by  $m_0$, and dipoles belonging to sublattice $\alpha:(f,af)$ have unit vector
$ \hat{ m}_i^\alpha = (\sin\theta_i^\alpha \cos\varphi^\alpha ,\sin\theta_i^\alpha\sin\varphi^\alpha ,\cos\theta_i^\alpha)$.

Because sublattices have easy planes mutually perpendicular, the full dipolar energy Eq.\ref{eq1} is separable into symmetric and antisymmetric long-range interactions (see Appendix\ref{sec:ana} for details).
They give rise to  four energetic contributions to the magnetic energy of the system and are consecutively denoted such that $\rm U_d=\frac{g}{2}\left[\sum_{i,k}\mathcal{U}_{ik}^{f}+\mathcal{U}_{ik}^{af}+\mathcal{U}_{ik}^{\rm J}+\mathcal{U}_{ik}^{\rm dm}\right]$. 
They correspond respectively to symmetric intra-sublattice interactions  in \emph{f}: $\mathcal{U}_{ik}^\emph{f}=J_{ik}^{0}\left(\frac{3}{2}\cos(\theta_i +\theta_k)-\frac{1}{2}\hat{m}_i\cdot\hat{m}_k\right)$ and  \emph{af}:  $\mathcal{U}_{ik}^\emph{af}=J_{ik}^{0}\left(\hat{m}_i\cdot\hat{m}_k\right)$, a symmetric inter-sublattice interaction: $\mathcal{U}_{ik}^{J}=J_{ik}\left(\hat{m}_i^\emph{f}\cdot\hat{m}_k^\emph{af}\right)$, and an antisymmetric inter-sublattice interaction energy $\mathcal{U}_{ik}^{\rm dm}=\bm{\mathcal{D}}_{ik}\cdot(\hat{m}_i^\emph{f}\times \hat{m}_k^\emph{af})$. The associated couplings between dipoles $i$ and $k$ read $J_{ik}^{0}=\frac{1}{|i-k|^3}$, $J_{ik}=\frac{1}{\left(\eta^2+(i-k+\frac{1}{2})^2\right)^{3/2}}$ which are respectively symmetric intra-chain and interchain couplings. $\bm{\mathcal{D}}_{ik}=-3\frac{\eta(i-k+\frac{1}{2})}{\left((i-k+\frac{1}{2})^2+\eta^2\right)^{\frac{5}{2}}})\hat{z}$ corresponds to an interchain Dzyaloshinskii–Moriya antisymmetric (DM) coupling \cite{dzyaloshinskii1958,moriya1960anisotropic,anderson1950antiferromagnetism}, perpendicular to the plane of the lattice \cite{mellado2022intrinsic}. 
\subsection{\label{sec:states}Equilibrium magnetic states}
 We have used molecular dynamics simulations to study the magnetic evolution of dipoles in the zig-zag lattice for different values of the offset $\eta$ (see Appendix \ref{sec:sim} for details). 
To examine the possible equilibrium magnetic configurations of the zig-zag lattice, we solved the equations of motion of each dipole when sublattice \emph{af} receded from \emph{f} (when $\eta$ grows) using molecular dynamics simulations. Details can be found in Appendix ~\ref{sec:sim}. In the simulations, each dipole $i$ consists of a magnetic bar with inertia moment $I$ and magnetic moment intensity $m_0$, which rotates with angle $\theta_i$ respect to its local axis of rotation. The evolution of $\theta_i$ is due to the torques $\mathcal{T}_i$ produced by dipolar interaction of dipole $i$ with all other dipoles in the lattice. The angular rotation has damping product of the friction at the rotation axis as shown by Eq.\ref{eq:dm}.
Depending on the offset $\eta$, dipoles settle into four magnetic configurations. At $\eta\in (0,0.2)$, the zig-zag lattice realizes an antiferromagnetic (AF) phase, along $\hat{z}$. At $\eta\in (0.2,0.6)$, all dipoles settle in the $\hat{x}-\hat{y}$ plane where each sublattice is in an  AF state. At $\eta\in (0.6,0.8)$, the \emph{f} sublattice realizes metastability and can be found either in the ferromagnetic (F) or the AF configuration. But once $\eta\geq 0.8$, it settles in the F state and remains in this configuration for all values of $\eta\geq 0.8$. 
Sublattice \emph{af}, on the other hand, settles in an AF configuration from $\eta>0.2$ \cite{mellado2022intrinsic}. Further, in the interval when $\eta\in (0.85,1)$, two Bloch domain walls (DW),  inhabit the antiferromagnetic chain. As $\eta$ grows, the solitons move from the edges of the chain toward the center of \emph{af} until they meet and disappear at $\eta\sim 1.5$. At this offset, \emph{af} returns to the original AF ground state along $\pm\hat{y}$. To make contact with the  recent experiments reported in a zig-zag lattice of dipoles  \cite{mellado2022intrinsic}, we have used the same experimental values for $I$, $\xi$ and $m_0$ in all simulations (Appendix ~\ref{sec:sim}).  The magnetic configurations found coincide with those reported in the experiments of \cite{mellado2022intrinsic}. 
Results from numerical simulations of the zig-zag lattice with $n=39$ dipoles in \emph{f} and $n-1=38$ in \emph{af}, for several values of $\eta$ are shown in Figs.\ref{f2} and \ref{f3}.  Consider Fig.\ref{f2}(a). Arrows represent the projection of the magnetic moment of the dipoles in $af$ in the plane $\hat{x}-\hat{y}$. Dipoles configure a Bloch DW when their magnetization points along $\pm\hat{z}$. In Fig.\ref{f2}(a), the consecutive tiny arrows of the same color close to the edges of \emph{af} in the two upper rows give them away. As $\eta$ increases from top to lower rows, the coupling $\mathcal{D}$ drops. This causes two effects: 1) the width of the DW (being inversely proportional to $ \mathcal {D}$ as will be shown in the next section), increases. Consequently, they spread, though slightly and at a very small rate which causes no big damage to their structure, before the two meet.
2) As $\eta$ increases, the solitons move from the edges toward the center of the lattice. Once they have met at the middle point, \emph{af} relaxes back to one of its AF ground states $\pm\hat{y}$ (lower row Fig.\ref{f2}(a)). Figs.\ref{f2}(b-c) show the $\hat{y}$ and $\hat{z}$ components of the  magnetization vector of each dipole $k$ of $af$ of Fig.\ref{f2}(a), $m_y^k=\sin{\theta_k}$, $m_z^k=\cos{\theta_k}$. As $\eta$ varies, the motion of the DW along the lattice determines the magnetization  of $af$ in the plane $\hat{y}-\hat{z}$. 
Fig.\ref{f3} rationalizes the previous results in terms of the inhomogeneous internal fields of the system, by showing the energy landscape that sublattice \emph{f} produces along \emph{af} (Appendix ~\ref{sec:sim}). From top to bottom $\eta=0.98, 1.30, 1.94$. In the upper row of Fig.\ref{f3}, the two energy minima at the edges of \emph{af} coincide with the positions of the solitons as shown by the inset. As $\eta$ grows, the potential wells become shallower and move toward the center of \emph{af} (middle row); this coincides with the new locations of the Bloch DW. The lower row corresponds to the energy landscape when \emph{af} has returned to its ground state at $\eta=1.94$.\\
In what follows, we focus on the \emph{af} sublattice in the soliton regime, which  occur for offsets $\eta\geq0.8$. In this regime, simulations show that \emph{af} is in the AF state, and hosts two Bloch DWs for a range of values of $\eta$. On the other hand, \emph{f} remains in the F state, with a constant intrachain energy which from now on we set to zero. 
\subsection{\label{sec:effective}Long ranged symmetric and anti-symmetric couplings.}
Consider the energy of \emph{af}
\begin{eqnarray}
U=\frac{g}{2}\sum_{i,k}\mathcal{U}_{ik}^{af}+\mathcal{U}_{ik}^{\rm J}+\mathcal{U}_{ik}^{\rm dm}=U^{af}+U^{\rm J}+U^{\rm dm}
\end{eqnarray}
where $U^{\rm J}=\frac{g}{2}\sum_{ik} J_{ik}(\hat{\bm m}_k^\emph{af}\cdot \hat{\bm m}_i^\emph{f})$, $U^{\rm dm}=\frac{g}{2}\sum_{ik}\mathcal{\bm D}_{ik}\cdot(\hat{\bm m}_k^\emph{af}\times \hat{\bm m}_i^\emph{f})$ are the interchain energies associated to the symmetric and antisymmetric interchain couplings and $U^{af}$ is the intrachain dipolar energy of \emph{af}. In the soliton regime, $\hat{\bm m}^\emph{f}=\pm(1,0,0)$, which allows to write the interchain energy as $U^{\rm J}=m_0\sum_{k\in\emph{af}}\hat{\bm m}_k^\emph{af}\cdot{\bm h}_{J}^k$ and $U^{\rm dm}=m_0\sum_{k\in\emph{af}}\hat{\bm m}_k^\emph{af}\cdot{\bm h}_{dm}^k$, where ${\bm h}_{J}^k$ and ${\bm h}_{dm}^k$ are the internal fields created by dipoles of sublattice \emph{f} at sites $k$ of \emph{af} (see Appendix\ref{sec:ana} for details).

 It is convenient to introduce two sublattices in the antiferromagnetic chain, with unit magnetizations denoted by $\textbf{m}_1$ and $\textbf{m}_2$. They give rise to the antiferromagnetic vector $\textbf{n}=\frac{\textbf{m}_1-\textbf{m}_2}{2}$ and the total magnetization $\textbf{m}=\textbf{m}_1+\textbf{m}_2$. These are subjected to the constraints, $|\textbf{m}_1|^2=|\textbf{m}_2|^2$ which cause  $\textbf{m}\cdot\textbf{n}=0$. If the magnetization is small $|\textbf{n}|^2\approx 1$ \cite{kosevich1990magnetic}. In a state of equilibrium as the one illustrated in Fig.\ref{f1}(a) $\textbf{m}_1=-\textbf{m}_2$. A continuum theory of the $af$ chain operates with the two slowly varying fields $\textbf{m}_1(x)$ and $\textbf{m}_2(x)$, the magnetization $\textbf{m}(x)$ and the dominant staggered magnetization $\textbf{n}(x)$. Now, $m_0=\gamma\mathcal{J}$ where $\mathcal{J}$ is the density of angular momentum for one sublattice, and $\gamma$ is the gyromagnetic ratio.  The staggered magnetization is written in terms of the angular variables of Fig.\ref{f1} as $\textbf{n}(x)=(\sin\theta(x)\cos\phi,\cos\theta(x),\sin\theta(x)\sin\phi)$. Due to the imposed constraints on the magnetization fields,  the uniform magnetization follows the dynamics of the staggered magnetization $\textbf{m}=\frac{\chi}{m_0^2}(\mathcal{J}\dot{\textbf{n}}\times \textbf{n}+m_0\textbf{n}\times(\textbf{h}_{\rm ef}\times\textbf{n}))$ where $\chi$ is the paramagnetic susceptibility and $\textbf{h}_{\rm ef}$ the  effective field on \emph{af}.
In the continuum limit, the couplings become $J^0(x)=\frac{1}{x^3}$, $J(x,y)=\frac{1}{\left(y^2+x^2\right)^{3/2}}$ and $\bm{\mathcal{D}}(x,y)=\frac{3xy}{\left(x^2+y^2\right)^{\frac{5}{2}}}\hat{z}$. Integrating out the x coordinate in the interchain couplings yields $J(y)=\frac{1-\frac{1}{\sqrt{4 y^2+1}}}{y^2}$ and  $\mathcal{D}(y)=\frac{8y}{(1+4y^2)^{3/2}}$. $J(y)$ and $\mathcal{D}(y)$ account for the influence of the ferromagnetic chain on \emph{af}. The associated internal fields,  product of symmetric and chiral dipolar interactions, are respectively $\textbf{h}_{\rm J}(y)=\frac{m_0}{\chi}J(y)\hat{x}$ and $\textbf{h}_{\rm dm}(y)=\frac{m_0}{\chi}\mathcal{D}(y)\hat{y}$.
The total effective field acting on \emph{af} due to dipoles in \emph{f} is $\textbf{h}_{\rm ef}=\textbf{h}_{\rm J}+\textbf{h}_{\rm dm}=\frac{m_0}{\chi}(J(y),\mathcal{D}(y),0)$
\section{\label{sec:lagrangian}Dynamics of solitons} 
The Bloch domain walls \cite{da2014observation,taherinejad2012bloch} found above are localised structures which propagates along \emph{af} with shape retention. Therefore they constitute solitons \cite{bar2006dynamics}. In this section we use a continuum field theory to examine how the internal fields $\textbf{h}_{\rm J}(y)$ and $\textbf{h}_{\rm dm}(y)$ stabilize and accelerate solitons in the \emph{af} chain.
\subsection{\label{sec:lag}AFM field theory} 
In the non-dissipative continuum limit, the dynamics of magnetization fields in \emph{af} is determined by the lagrangian density \cite{dasgupta2017gauge,cuevas2014sine,benfatto2006derivation,dasgupta2018energy}
\begin{equation}
{\mathcal L}(\textbf{n}) = \frac{\rho}{2}|\dot{\textbf{n}}|^2-\rho\gamma \textbf{h}_{\rm ef}\cdot(\textbf{n}\times\dot{\textbf{n}})-\frac{\mathcal{A}}{2}|\textbf{n}^\prime|^2-\frac{\rho}{2}|\gamma \textbf{h}_{\rm ef}\times \textbf{n}|^2
\label{eq:L}
\end{equation}
The term $\frac{\rho}{2}|\dot{\textbf{n}}|^2$ is the kinetic energy of the staggered magnetization and $\rho=\chi/\gamma^2$ is the density of inertia of $\textbf{n}$. The second term $\rho\gamma \textbf{h}_{\rm ef}\cdot(\textbf{n}\times\dot{\textbf{n}})$  also known as gyroscopic \cite{gonzalez2022gyroscopic} quantifies the effective geometric phase for the dynamics of $\textbf{n}$, due to $\textbf{h}_{\rm ef}$ \cite{dasgupta2018energy}. The two last terms correspond to the potential energy. In the context of the zig-zag chain studied here, the first term is proportional to the exchange strength $\mathcal{A}=\frac{m_0}{\gamma^2h_{ex}}>0$ where the exchange field $h_{ex}$ is proportional to the energy of the uniform antiferromagnetic ground state. This term is due to the symmetric intrachain interactions in $af$ which favor a uniform antiferromagnetic state ($\partial_x\textbf{n}=0$). The potential term $\frac{\rho}{2}|\gamma \textbf{h}_{\rm ef}\times \textbf{n}|^2\sim\frac{m_0^2}{2\chi}(\mathcal{D}^2\sin{\theta}^2)$  expresses local induced uniaxial anisotropy favoring the direction of the effective field.  For the case at hand it is due to the DM interaction emerging from the dipolar interaction in the system and can be written as $\frac{K}{2}|\hat{y}\times \textbf{n}|^2$ where $K=\frac{\mathcal{D}^2m_0^2}{\chi}$. 

Eq.\ref{eq:L} can be recast as 
${\mathcal L}(\textbf{n}) = \frac{\rho}{2}|\dot{\textbf{n}}|^2-\rho\gamma \textbf{h}_{\rm ef}\cdot(\textbf{n}\times\dot{\textbf{n}})-\mathcal{U}(\textbf{n})
$, where the potential energy density reads 
\begin{eqnarray}
\mathcal{U}(\textbf{n})=\frac{\mathcal{A}}{2}\abs[\Big]{\frac{\partial\textbf{n}}{\partial x}}^2+\frac{K}{2}|\hat{y}\times \textbf{n}|^2 
\label{eq:Un}
\end{eqnarray}
\emph{af} has two uniform ground states $\textbf{n}=\pm\hat{y}$, excitations manifest as linear spin waves and non-linear solitons. Here we focus in the solitons which take the form of Bloch DWs. They have associated a  width $\lambda=\sqrt{\mathcal{A}/{K}}$, which defines a characteristic length scale of the field theoretical problem. The characteristic scales of time and energy are given respectively by  $\tau=\sqrt{\frac{\rho}{K}}$ and $\epsilon=\sqrt{\mathcal{A}K}$, where $\epsilon$ is the energy density of the DW \cite{belashchenko2016magnetoelectric}. 
\subsection{\label{sec:newton}Electromagnetic force on the soliton}
The dynamics of $\textbf{n}$ is obtained from the Euler Lagrange equations from the lagrangian density Eq.\ref{eq:L}. In the cartesian coordinate system shown in Fig.\ref{f1}, axis $\hat{x}$ is perpendicular to the DW and axis $\hat{z}$ parallel to it. For the analysis of the solitons in Fig.\ref{f2}, it is convenient to use the language of collective coordinates \cite{tchernyshyov2022unified} and parametrize them by the variables $q_1\equiv X$ and $q_2\equiv\Phi$, corresponding to their position along the lattice and the azimuthal angle, respectively \cite{bouzidi1990motion}. Thus, the DW profile can be written as: 
\begin{equation}
\cos\theta(x)=\pm \tanh\frac{x-X}{\lambda}, \; \phi(x)=\Phi
\label{eq:dw}
\end{equation}
Eq.\ref{eq:dw} defines a static DW and minimizes Eq.\ref{eq:Un}, where $\theta$ and $\phi$ are the polar and azimuthal angles parameterizing the unit vector $\textbf{n}$ \cite{kosevich1990magnetic}. To describe the low energy dynamics of the domain wall, we promote the two collective coordinates to dynamics variables in the DW anzats Eq.\ref{eq:dw}. The variation of $\textbf{n}=(\sech\frac{x-X}{\lambda}\cos\phi,\tanh\frac{x-X}{\lambda},\sech\frac{x-X}{\lambda}\sin\phi)$ in time is mediated by the change of these collective coordinates: $\dot{\textbf{n}}=\dot{q}_i\partial\textbf{n}/\partial q_i=\dot{X}\partial\textbf{n}/\partial X$ where repeated indices are implicitly summed over and we used that $\dot{\Phi}=0$. The kinetic energy of the DW  becomes $M_{ij}\dot{q_i}\dot{q_j}/2$, where the inertia tensor is defined as $M_{ij}=\rho \int(\frac{\partial \textbf{n}}{\partial q_i}\cdot\frac{\partial \textbf{n}}{\partial q_j})dx$ \cite{dasgupta2017gauge}. For the case at hand, $M_{\Phi\Phi}=2\rho\lambda\equiv I, M_{XX}=2\rho/\lambda\equiv M, M_{X\Phi}=0$, where $M$ and $I$ are the mass and moment of inertia per unit length of the DW.The kinetic energy of the DW becomes $\frac{1}{2}(I\dot{\Phi}^2+M\dot{X}^2)=\frac{\rho}{\lambda}\dot{X}^2$ (Appendix\ref{sec:ana}).   

In the Lagrangian of the soliton, the gyroscopic term \cite{thiele1973steady} can be written in terms of a gauge field as $A_i\dot{q}_i$ where $A_i(X,\Phi)=\rho\gamma\int {\rm\textbf{h}}_{\rm ef}\cdot(\frac{\partial \textbf{n}}{\partial q_i}\times \textbf{n})dx$ \cite{dasgupta2017gauge}. The gauge potential for the domain wall of Eq.\ref{eq:dw}  has components that depend on $y$ (Appendix\ref{sec:ana}):
\begin{equation}
\begin{aligned}
A_X(y) &=\pm\frac{\pi \rho\gamma }{2}\rm h_{\rm J}\sin\Phi \\
A_\Phi(y) =&\pm\rho\gamma \lambda(\rm h_{\rm dm}-\rm h_{\rm J}\cos\Phi)
\end{aligned}
\end{equation}
The curl of A yields an emergent magnetic field \cite{dasgupta2017gauge}, 
\begin{eqnarray}B_{X\Phi}(y)=\pm\rho\gamma(\rm h_{\rm dm}-\frac{\pi}{2}\rm h_{\rm J}\cos\Phi)
\label{B}
\end{eqnarray}
This field, the product of the interchain dipolar coupling,  decays with $y$. When $\rm{B}_{X\Phi}$ is time-dependent, it induces an electric field $E_i=\int\rho\gamma \dot{\textbf{h}}_{\rm ef}\cdot(\frac{\partial \textbf{n}}{\partial q_i}\times \textbf{n})dx$ \cite{tchernyshyov2022unified}. 

In the experiment of \cite{mellado2022intrinsic}, the sublattices of the zig-zag lattice move apart at speed $v=\frac{\partial y}{\partial t}$. In this case, the effective magnetic field becomes dependent on time $\dot{\textbf{h}}_{\rm ef}=\frac{\partial\textbf{h}_{ef}}{\partial y} v$ and gives rise to an emergent electric field with components $E_{X}(y)=\pm\rho\gamma v\frac{\pi}{2} \sin\Phi\frac{\partial\rm{h}_{\rm J}}{\partial y}$ and $E_{\Phi}(y)=\pm\lambda\rho\gamma v\frac{\partial\textbf{h}_{\rm dm}}{\partial y}$.
The emergent fields satisfy Jacobi identities and hence Maxwell equations $\bf{\nabla}\times\textbf{ E}+\dot{\textbf{B}}=0$ and $\bf{\nabla}\cdot\textbf{B}=0$ \cite{dasgupta2018energy,tchernyshyov2022unified}.
The previous numerical analysis indicate that solitons are rigid, thus they have only one continuous degrees of freedom which is the zero mode $X$ associated to the global translational symmetry along an infinite \emph{af}. The effective field $h_{ef}(y)$ respects that symmetry. Therefore for the solitons of Fig.\ref{f2} where $\phi=\pi/2$ we have $\dot{\phi}=0$ and $B_{X\Phi}=\pm\rho\gamma\rm h_{\rm dm}$. The electromagnetic force on the DW along the lattice axis ($\hat{x}$), is such that $F_{X}=E_X=\pm\rho\gamma v\frac{\pi}{2}\frac{\partial\rm{h}_{\rm J}}{\partial y}=-\frac{\partial A_X}{\partial t}$ and depends only on the emergent electric field. Consequently, the implicit time dependence of $h_J(y)$ allows the acceleration of the DW along the zig-zag lattice as long as there is a relative velocity between its sublattices (Appendix\ref{sec:ana}). 
\section{\label{sec:conclusion}Summary and Discussion}
 Finding new strategies to propel antiferromagnetic domain walls remains a challenge. The purpose of this work has been to examine whether and under which circumstances stable antiferromagnetic domain walls could be driven by internal emergent fields arising at magnetic interfaces. To that aim, we modeled an interface of two magnetic lattices by a zig-zag chain of easy plane dipoles interacting via dipolar coupling. We found that such a model realizes a planar magnetic state with one sublattice antiferromagnetic  and the other in the ferromagnetic state. The internal Dzyaloshinskii-Moriya field, which arises from the inter-sublattice dipolar interactions, stabilizes chiral solitons at each edge of the antiferromagnetic sublattice.  The dynamics of such solitons is studied by deriving the long-wavelength lagrangian density of this non-conventional easy axis antiferromagnetic chain in the soliton regime. Using the collective coordinates formalism, we find that aside from conservative forces, the internal fields due to the chiral and symmetric couplings between sublattices generate an effective magnetic field exerting a gyrotropic force on the solitons and an induced electric field when the two chains move apart along the lattice inversion axis. The emergent electromagnetic fields satisfy Maxwell equations and produce a Lorentz force that can accelerate solitons. As a result, from the edges of $af$, two Bloch DW migrate to the center of the lattice before they meet and disappear. 
An important aspect of this work is that, once the magnetization vectors are plugged into the formula for the dipolar energy, it is possible to separate antisymmetric and symmetric interchain dipolar contributions. This step allows to track the role of each of them in the subsequent continuum analysis of the soliton regime: while the antisymmetric contribution gives rise to an internal magnetic field able to produce Bloch DWs at the edge of the lattice, once the solitons have arisen, it is the symmetric contribution to the energy which produces the force  that accelerates the solitons along the lattice as long as the gap between sublattices changes with time. This means that a possible strategy to accelerate solitons could involve magnetic fields that point along the lattice but change in time orthogonal to it.
 
  A natural next step could generalize this model into two dimensions. That would emulate a more realistic version of the magnetic interface and allow us to test whether the unconventional dynamics shown here remain when the two chains are extended in the plane $\hat{x}-\hat{y}$.  Another exciting venue to be studied is the effect of an electric current passing through the interface. When the spin of electrons encounters a magnetic texture undergoes precession and exchange angular momentum with the soliton \cite{tchernyshyov2022unified}. While it has been shown \cite{dasgupta2017gauge} that the adiabatic spin torque cannot propel a DW on its own, it would be interesting to examine its effect in the model presented here.
\begin{acknowledgments}
P.M. and I.T acknowledge support from Fondecyt under Grant No. 1210083.
\end{acknowledgments}
\appendix
\section{\label{sec:sim}Molecular dynamics simulations}
The discrete magnetic dipoles are modeled as mechanical rods. The orientation of each dipole is characterized by an unitary vector ${\bf m}_i = (\sin{\theta_i} \cos{\phi_i}, \sin{\theta_i} \sin{\phi_i}, \cos{\theta_i})$. The angular variable $\phi_i$ is fixed and defines a plane of rotation with normal vector ${\bf p}_i = (-\sin{\phi}, \cos{\phi}, 0)$. The angular variable $\theta_i$ is a dynamical variable that changes in time due to the interaction of the magnetic dipole $i$ with its environment.
The equation of motion for the polar angles of inertial dipoles located at sites $i$ in sublattice $\alpha$, interacting through the full long-range dipolar potential with all other dipoles in the lattice, reads:
\begin{equation}
I\frac{d^2 \theta_{i}^\alpha}{dt^2}=\mathcal{T}_i^\alpha -\xi \frac{d \theta_{i}^\alpha}{dt}  
\label{eq:dm}
\end{equation}
 Where $I$ denotes the moment of inertia of the magnets, $\xi$ is the damping for the rotation of dipoles in the lattice. The first term at the right hand side of the previous equation is the magnetic torque due to the action of the internal magnetic field from dipolar interaction between all dipoles in the system $\mathcal{T}_i^\alpha=\hat{\bf m}_{i}^\alpha\times \rm{\bf h}_i^\alpha$, where $\rm {\bf h}_i^\alpha=\frac{\partial U_d}{\partial \hat{\bf m}_i^\alpha}$ denotes the internal magnetic field produced by all dipoles  but the $i$-$th$ at the position of $\hat{\bf m}_{i}^\alpha$ and the potential energy defining the dipolar magnetic interactions is given by Eq.\ref{eq1}.  
 Following previous experiments we used in the simulations dipolar rods with radius $r=0.79\times 10^{-3}$ m, length $L=12.7\times 10^{-3}$ m, and mass $=0.19\times 10^{-3}$ kg, saturation magnetization $M_s=1.05\times10^{6}$ $\rm A/m$, inertial moment $I=1.53\times 10^{-9}$ $\rm kg m^2$ and magnetic moment $m_0=0.026\times10^{6}$ $\rm Am^2$.
The second term on the RHS of Eq.\ref{eq:dm} corresponds to a damping term. The viscosity $\xi(\theta_i)$ consists of a static and dynamics contribution. Here we used experimental values reported in \cite{mellado2022intrinsic}. The static part the damping was computed directly by fitting the relaxation of a single rod under a perturbation to the solution $\theta\sim\exp{-t/\tau_D}$. The estimated damping time $\tau_D\sim I/\xi$ of a single rod to be $\tau_D=0.83$ s. The dynamics component is due to the rotation of dipoles in the lattice with other magnets and therefore is dependent on the orientation of the dipole, being maximum when it is orientated perpendicular to $\hat{\bf z}$ axis. \\
We solved this set of coupled equations using a discretized scheme for the integration of the differential equation, given by the recursion
\begin{equation} 
\theta_i(t + \Delta t) = f_a \theta_i(t) - f_b \theta_i(t - \Delta t) + f_c \left( {\bf p}_i \cdot \mathcal{T}_i(t) \right) \, \text{,}
\end{equation}
where the functions $f_a$, $f_b$ and $f_c$ corresponds to
\begin{equation}
f_a = \frac{2}{1 + \Delta t \, \eta(\theta_i(t)) / 2} \, \text{,}
\end{equation}
\begin{equation}
f_b = \frac{1 - \Delta t \, \eta(\theta_i(t))/2}{1 + \Delta t\,  \eta(\theta_i(t)) / 2} \, \text{,}
\end{equation}
\begin{equation}
f_c = \frac{\Delta t^2}{I(1 + \Delta t \, \eta(\theta_i(t)) / 2)} \, \text{.}
\end{equation}
We performed the simulations for systems with a total of $2n-1=77,97$ and $117$ magnetic dipoles, starting with the sublattices close together and separating \emph{af} from \emph{f} at constant speed. The fixed angular variable $\phi$ that defines the plane of rotation was chosen as $0$ for \emph{f}n and $\pi/2 + \delta \phi_i$ for \emph{af}. The quantity $\delta \phi_i$ corresponds to a Gaussian random noise with zero mean and standard deviation equal to $0.01$.

The time interval used to integrate the equations of motion was $\Delta t = 2 \times 10^{-6}$, and at each time step, sublattice \emph{af} was moved apart from \emph{f} along the $\hat{y}$ direction by a distance of  $\delta\eta=3.5\times 10^{-8}$, or $a_y=1.6\times 10^{-6}a_x$, with $a_x= 2.2\times 10^{-2}$ the distance between neighbour magnetic dipoles in the same sublattice. This integration time step is smaller than the shortest time scale in the system which can be estimated from the dipolar force between two nearest neighbor dipoles, and scales as $\tau_c\sim8\pi\sqrt{\frac{\pi I L}{\mu_0}}\frac{a_x}{m_0}\sim0.2$ s. Therefore, at every time step of the integration, dipoles have relaxed to their equilibrium angular positions.\\
To examine the soliton regime ($\eta>0.8$) the initial distance between sublattices was set to $\eta=0.5$. Initial conditions for angular positions of dipoles in the the zig-zag lattice were antiferromagnetic for both sublattices, which is the stable magnetic configuration at $\eta\sim 0.5$, and the transition to the solitonic regimen was seen at $\eta\sim 1$. As $\eta$ increased from $0.5$ to $2$ we stored the angular variable $\theta_i$ for each dipole $i$ at every time step.\\
Stability was also studied, by suddenly stopping the relative movement between the chains in the soliton regime and then waiting a time of $\sim2$s to observe the evolution of the angles. Solitons were found to be stable for distances near the transition, with small oscillations that did not destroy the profile of the chiral structures.\\
In Fig.\ref{f2}, we show the magnetization of each dipole in $af$ chain for different distances along the simulation in the soliton regimen. Fig.\ref{f2}(a) illustrates the configuration of the dipoles as seen from a plane perpendicular to $\hat{x}-\hat{z}$ plane. This corresponds to the $y$ component of the magnetization of each dipole. The colors are interchanged every time there is a change of sign on the $y$ component of the magnetization to illustrate the antiferromagnetic state. This color change is broken at the location of the domain walls that corresponds with the position of the soliton. Fig.\ref{f2}(b) and Fig.\ref{f2}(c) show the projections of the magnetization along the $\hat{y}$ and $\hat{z}$ axis respectively. 
 The energy landscape shown in Fig.\ref{f3} corresponds to the effective torque $\mathcal{T}$ that produces chain $f$ over each magnet of chain $af$ projected on the plane perpendicular to the magnet affected by said torque. This is computed as $\rm (\mathcal{T}_i)_{\parallel} = \left \vert {\bf \mathcal{T}}_i - \left( {\bf \hat{m}}_i \cdot {\bf \mathcal{T}}_i  \right) {\bf \hat{m} }_i \right \vert$.
 \section{\label{sec:ana}Calculations details}
 \subsection{Energy separation and magnetic couplings}
The dimensionless dipolar energy of the system can be written as: 
\begin{equation}
\rm U_{\rm d}=(U^f+U^{af}+U^{\rm J}+U^{\rm dm}) \, \text{,}
\label{eq:Energy2}
\end{equation}
where $\rm U^{af}$ is the dipolar interaction between magnets belonging to sublattice \emph{af}, $U^{f}$ is the dipolar interaction between magnets belonging to sublattice \emph{f}, and $\rm U^{\rm J}+U^{\rm dm}$ is the interaction between magnets belonging to distinct sublattices.  
After replacement of the magnetization vectors for each sublattice, in terms of angular variables these energy contributions become: 
\begin{eqnarray}
\rm U^{f}&=&\frac{1}{2^3} \sum_{i\neq k} \frac{1}{|i-k|^3} \left[  \cos(\theta_i^f-\theta_k^f) -3  \sin\theta_i^f  \sin \theta_k^f   \right]\nonumber\\
\rm U^{af}&=&\frac{1}{2^3} \sum_{i\neq k} \frac{1}{|i-k|^3} \left[  \cos(\theta_i^{af} -\theta_k^{af})  \right]\nonumber\\
\rm U^{\rm J}+U^{\rm dm}&=&\sum_{i\neq k}\left[ \frac{1}{\left(\eta^2+(i-k)^2\right)^{3/2}}\left( \cos\theta_i^f\cos\theta_k^{af} -3 \frac{\eta(i-k+\frac{1}{2})}{(i-k+\frac{1}{2})^2+\eta^2}\sin\theta_i^f\sin\theta_k^{af} \right)\right]\nonumber\\ \nonumber\\
\label{eq:EnergyAngles}
\end{eqnarray}
where sub-indices $i$ and $k$ index the positions of dipoles in the zig-zag lattice such that $|i-k|$ is the distance between them. The magnetic moment of dipole $i$($k$) at sublattice $f$ ($af$) is denoted by $\hat{m}_i^f$ ($\hat{m}_k^{af}$),  and $\theta_i^f$ ($\theta_k^{af}$) denotes the polar angle of dipole $i$ ($k$) in sublattice $f$ ($af$).

The previous expressions can be written as,
\begin{eqnarray}
\rm U^{f}& = & \frac{1}{8}\sum_{i\neq k} \frac{1}{|i-k|^3} \left[-\frac{1}{2}\hat{m}_i^f\cdot\hat{m}_k^f +\frac{3}{2}\cos(\theta_i^f +\alpha_k^f) \right]\nonumber\\
U^{af}& = & \frac{1}{8}\sum_{i\neq k} \frac{1}{|i-k|^3} \left[\hat{m}_i^{af}\cdot\hat{m}_k^{af} \right]\nonumber\\
\end{eqnarray}
\begin{eqnarray}
\rm U^{\rm J}+U^{\rm dm}&=&\sum_{i\neq k}\left[ \frac{1}{\left(\eta^2+(i-k+\frac{1}{2})^2\right)^{3/2}}\left(\cos\theta_i^f\cos\theta_k^{af}-3 \frac{\eta(i-k+\frac{1}{2})}{(i-k+\frac{1}{2})^2+\eta^2}\sin\theta_i^f\sin\theta_k^{af} \right)\right]\nonumber\\
\end{eqnarray}
\begin{eqnarray}
\rm U^{\rm J}+U^{\rm dm}&=&\sum_{i\neq k}\left[ \frac{1}{\left(\eta^2+(i-k+\frac{1}{2})^2\right)^{3/2}}\left(\frac{1}{2}\hat{m}_i^f\cdot\hat{m}_k^{af}-3 \frac{\eta(i-k+\frac{1}{2})}{(i-k+\frac{1}{2})^2+\eta^2}\hat{z}\cdot(\hat{m}_i^f\times \hat{m}_k^{af}) \right)\right]\nonumber\\
\end{eqnarray}
And the dipolar energy becomes:
\begin{eqnarray}
\rm U_d=\sum_{i\neq k}\left[J_{ik}^0\left(-\frac{1}{2}\hat{m}_i^f\cdot\hat{m}_k^f +\frac{3}{2}\cos(\theta_i^f +\theta_k^f) +\hat{m}_i^{af}\cdot\hat{m}_k^{af}\right)+J_{i,k}\left(\hat{m}_i^f\cdot\hat{m}_k^{af}\right) +\bm{\mathcal{D}}_{ik}\cdot(\hat{m}_i^f\times \hat{m}_k^{af})\right]\nonumber\\
\label{energyfin}
\end{eqnarray}
where $$J^0_{ik}=\frac{1}{|i-k|^3}$$ and $$J_{ik}=\frac{1}{\left(\eta^2+(i-k+\frac{1}{2})^2\right)^{3/2}}$$ are respectively  the exchange coupling between dipoles belonging to same  and different chains. 
The third term in the right hand side of Eq. \ref{energyfin}  is a Dzyaloshinskii–Moriya (DM) antisymmetric type of exchange perpendicular to the plane of the system, 
$$\bm{\mathcal{D}}_{ik}=\left(0,0,-3\frac{\eta(i-k+\frac{1}{2})}{\left((i-k+\frac{1}{2})^2+\eta^2\right)^{\frac{5}{2}}}\right) \, \text{.}$$
 \subsection{Internal fields in the soliton regime}
Consider the energy from the antisymmetric interchain coupling 
\begin{equation}
\begin{split}
    U^{\rm dm}& =\frac{g}{2}\sum_{ik}\mathcal{\bm D}_{ik}\cdot(\hat{\bm m}_k^\emph{af}\times \hat{\bm m}_i^\emph{f})=\frac{g}{2}\sum_{k\in af}\hat{\bm m}_k^\emph{af}\cdot\sum_{i\in f}(\mathcal{\bm D}_{ik}\times \hat{\bm m}_i^\emph{f})\\
    &=\pm\frac{g}{2}\sum_{k\in af}\hat{\bm m}_k^\emph{af}\cdot\sum_{i\in f}\mathcal{\bm D}_{ik}(\hat{z}\times \hat{x})=m_0\sum_{k\in af}\hat{\bm m}_k^\emph{af}\cdot{\bm h}_{\rm dm}^k
 \end{split}
\end{equation}
with ${\bm h}_{\rm dm}^k=\pm\frac{g}{2m_0}\sum_{i\in f}\mathcal{D}_{ik}\hat{y}$, and where we used that in the soliton regime $\hat{\bm m}^{af}=\pm(1,0,0)$.

Similarly, the energy from the symmetric interchain coupling yields,
\begin{equation}
U^{\rm J}=\frac{g}{2}\sum_{ik} J_{ik}(\hat{\bm m}_k^\emph{af}\cdot \hat{\bm m}_i^\emph{f})=\frac{g}{2}\sum_{k\in af}\hat{\bm m}_k^\emph{af}\cdot\sum_{i\in f}J_{ik}\hat{\bm m}_i^\emph{f}
=m_0\sum_{k\in af}\hat{\bm m}_k^\emph{af}\cdot{\bm h_{\rm J}}^k
\end{equation}
where ${\bm h}_{\rm J}^k=\pm\frac{g}{2m_0}\sum_{i\in f}J_{ik}\hat{x}$
 \subsection{Inertia Tensor}
The kinetic energy of the DW  is given by $M_{ij}\dot{q_i}\dot{q_j}/2$, where the inertia tensor is defined as $M_{ij}=\rho \int(\frac{\partial \textbf{n}}{\partial q_i}\cdot\frac{\partial \textbf{n}}{\partial q_j})dx$. In our case, $ M_{X\Phi}=0$ and
 \begin{equation}
 M_{XX}=2\rho\lim_{L\to\infty}\int_X^L\left(\frac{\partial n}{\partial X}\cdot\frac{\partial n}{\partial X}\right)\,dx=\lim_{L\to\infty}2\rho\frac{\tanh{\frac{(x-L)}{\lambda}}}{\lambda}=\frac{2\rho}{\lambda}    
 \end{equation}
  \begin{equation}
 M_{\Phi\Phi}=\rho\lim_{L\to\infty}\int_X^L\left(\frac{\partial n}{\partial \Phi}\cdot\frac{\partial n}{\partial \Phi}\right)\,dx=\lim_{L\to\infty}\rho\lambda\tanh{\left(\frac{x-L}{\lambda}\right)}=2\rho\lambda    
 \end{equation}
 \subsection{Gauge Potential $A_X$ and $A_\phi$}
The gauge potential for the domain wall of Eq.\ref{eq:dw} is defined as $A_i(X,\Phi)=\rho\gamma\int {\rm\textbf{h}}_{\rm ef}\cdot(\frac{\partial \textbf{n}}{\partial q_i}\times \textbf{n})dx$. Its components depend on $y$,
 \begin{equation}
 \begin{split}
 A_X&=\rho\gamma\lim_{L\to\infty}\int_X^L(\textbf{h}_{\rm J}+\textbf{h}_{\rm dm})\cdot\left(\frac{\partial n}{\partial X}\times n\right)\,dx=\pm\rho\gamma\lim_{L\to\infty}\int_X^L\frac{\textbf{h}_{\rm J}\sech{(\frac{x-X}{\lambda})}\sin{\Phi}}{\lambda}\,dx\\
 & = \pm\frac{\pi}{2}\rho\gamma\textbf{h}_{\rm J}\sin{\Phi}  
 \end{split}
 \end{equation}
  \begin{equation}
 \begin{split}
 A_\Phi& =\rho\gamma\lim_{L\to\infty}\int_X^L(\textbf{h}_{\rm J}+\textbf{h}_{\rm dm})\cdot\left(\frac{\partial n}{\partial \phi}\times n\right)\,dx \\
 &=\pm\rho\gamma\lim_{L\to\infty}\int_X^L\sech{\left(\frac{x-X}{\lambda}\right)}\left[\textbf{h}_{\rm dm}\sech{\left(\frac{x-X}{\lambda}\right)}-\textbf{h}_{\rm J}\cos{\Phi}\tanh{\left(\frac{x-X}{\lambda}\right)}\right]\,dx\\
 & =\pm\rho\gamma\lambda\lim_{L\to\infty}\left[\textbf{h}_{\rm J}\cos{\Phi}\tanh{\left(\frac{L-X}{\lambda}\right)}+\textbf{h}_{\rm dm}\tanh{\left(\frac{L-X}{\lambda}\right)}\right]\\
 & =\pm\rho\gamma\lambda(\textbf{h}_{\rm dm}-\textbf{h}_{\rm J}\cos{\Phi})
 \end{split}
 \end{equation}
 \subsection{Emergent  magnetic field $B_{X\Phi}$}
 The curl of the vector potential  yields an emergent magnetic field product of the interchain dipolar coupling,
  \begin{equation}
 \begin{split}
 B_{X\Phi}&
 =\frac{\partial A_\Phi}{\partial X}-\frac{\partial A_X}{\partial \Phi}=\pm 2\rho\gamma\lim_{L\to\infty}\int_X^L(\textbf{h}_{\rm J}+\textbf{h}_{\rm dm})\cdot\left(\frac{\partial n}{\partial X}\times \frac{\partial n}{\partial\Phi}\right)\,dx\\
& =\pm\frac{2\rho\gamma}{\lambda}\lim_{L\to\infty}\int_X^L\left[\sech{\left(\frac{x-X}{\lambda}\right)^3}(\textbf{h}_{\rm J}\cos{\Phi}+\textbf{h}_{\rm dm}\sinh{\left(\frac{x-X}{\lambda}\right)}\right]\,dx\\
& =\pm 2\rho\gamma\lim_{L\to\infty}\left[\textbf{h}_{\rm dm}\tanh{\left(\frac{L-X}{\lambda}\right)^2}+\textbf{h}_{\rm J}\cos{\Phi}F(L)\right]\\
&=\pm\rho\gamma(\textbf{h}_{\rm dm}-\frac{\pi}{2}\textbf{h}_{\rm J}\cos{\Phi})
\end{split}
\end{equation}
where $F(L)=\left(\arctan\left(\sinh{\left(\frac{L-X}{\lambda}\right)}\right)+\sech{\left(\frac{L-X}{\lambda}\right)}\tanh{\left(\frac{L-X}{\lambda}\right)}\right)$
\subsection{Emergent  electric fields}
Since the sublattices of the zig-zag chain move apart, the effective magnetic field becomes dependent on time and gives rise to an emergent electric field,
\begin{equation}
\begin{split}
 E_X&=-\rho\gamma\lim_{L\to\infty}\int_X^L \dot{\textbf{h}}_{\rm ef}\cdot\left(\frac{\partial n}{\partial X}\times n\right)\\
 & =\pm\frac{\rho\gamma v}{\lambda}\partial_y\textbf{h}_{\rm J}\lim_{L\to\infty}\int_X^L \sech{\left(\frac{x-X}{\lambda}\right)}\sin{\Phi}\\
 & =\pm\rho\gamma v\partial_y\textbf{h}_{\rm J}\lim_{L\to\infty}\arctan{\left(\sinh{\left(\frac{L-X}{\lambda}\right)}\right)}\sin{\Phi}\\
  & =\pm\frac{\pi\rho\gamma v}{2}\partial_y\textbf{h}_{\rm J}\\
    & =\pm\frac{\pi\rho\gamma v}{\eta^3(1+4\eta^2)^{3/2}}\left[-1+\sqrt{1+4\eta^2}+\eta^2(-6+4\sqrt{1+4\eta^2})\right]
 \end{split}
\end{equation}
\begin{equation}
\begin{split}
E_\Phi &=-\rho\gamma\lim_{L\to\infty}\int_X^L \dot{\textbf{h}}_{\rm ef}\cdot\left(\frac{\partial n}{\partial X}\times n\right)\\
&= \pm\rho\gamma v\lim_{L\to\infty}\int_X^L \sech{\left(\frac{x-X}{\lambda}\right)}\left(\partial_y\textbf{h}_{\rm dm}\sech{\left(\frac{x-X}{\lambda}\right)}-\partial_y\textbf{h}_{\rm J}\cos{\Phi}\tanh{\left(\frac{x-X}{\lambda}\right)}\right)\\
&=\pm\rho\gamma v\lambda\partial_y\textbf{h}_{\rm dm}\\
&= \pm\rho\gamma v\lambda\frac{8(8\eta^2-1)}{(1+4\eta^2)^{5/2}}
 \end{split}
\end{equation}
 \subsection{Lorentz Force}
 The force on the soliton along the x direction is given by the electric field along X and the gyroscopic force, analogous of the Lorentz force \cite{dasgupta2018energy}
 \begin{equation}
 F_X=E_X+B_{X\Phi}\dot{\Phi}=E_X
\end{equation}
Using the previous results is easy to find that the time derivative of the gauge potential $\frac{dA_X}{dt}=-E_X$, and therefore $ F_X=-\frac{dA_X}{dt}$
 \bibliography{ferro}
\end{document}